\documentclass[
 reprint,
 superscriptaddress,
 amsmath,amssymb,
 aps,
 prb,
 longbibliography
]{revtex4-2}

\usepackage{graphicx}      
\usepackage{dcolumn}       
\usepackage{bbm}
\usepackage{bm}            
\usepackage[colorlinks=true, allcolors=blue]{hyperref} 

\DeclareMathOperator{\Tr}{Tr}

\DeclareMathOperator{\Real}{Re}
\DeclareMathOperator{\Imag}{Im}

\newcommand{\mycomment}[1]{}

\begin{document}

\title{Electron state tomography from quasiparticle interference maps}

\author{A. Razanajatovo}
\author{J. Cayssol}
\author{C. Dutreix}
\email{clement.dutreix@u-bordeaux.fr}

\affiliation{Univ. Bordeaux, CNRS, LOMA, UMR 5798, F-33400 Talence, France}

\date{\today}

\begin{abstract}
Characterizing electronic band structures requires precise knowledge of wave functions and their quantum geometry.
Here, we introduce a tomography method to reconstruct the density matrix of electron states from quasiparticle interference maps around single impurities.
We consider two-orbital models on a honeycomb lattice, relevant to graphene heterostructures and direct-gap semiconductors.
For on-site impurities, backscattering between time-reversed states directly maps the density matrix populations and coherences into distinct orbital contributions in the interference map.
While local probes usually lack orbital resolution, these orbital contributions transform under distinct symmetry group representations and can thus be disentangled to reveal the density matrix and quantum geometric tensor of the scattering states.
This establishes impurities as tomographic probes for band structures in scanning tunneling microscopy using conventional, unpolarized tips.
%
\end{abstract}

\maketitle


\textit{Introduction} --- Quantum geometry, captured by the quantum geometric tensor (QGT), plays a fundamental role in a wide range of electronic phenomena in solids.
Its imaginary part, relating to the Berry curvature, governs the geometric phase of wave functions~\cite{berry1984,Zak1989}. It is essential to understand properties as various as polarization~\cite{Vanderbilt1993,Resta1994}, orbital magnetism~\cite{Raoux2014,Raoux2015}, adiabatic pumping~\cite{Thouless1983}, electron dynamics under electromagnetic fields~\cite{Xiao2010}, quantum Hall effects~\cite{Klitzing1980,Thouless1982,Haldane1988,Jungwirth2002,novoselov2005,Zhang2005,Kane2005,Bernevig2006,Konig2007,Veyrat2020}, and to classify topological states~\cite{Fu2011,Po2018,Schindler2018HOTI,Schindler2018Bismuth,Bradlyn2017}. The QGT real part, known as the (Fubini-Study) quantum metric, quantifies the distance between nearby states in parameter space~\cite{bures1969,provost1980}. Though less explored, it has recently been shown to affect the localization of Wannier orbitals and supercurrents in flat-band superconductors~\cite{Peotta_2015,PhysRevLett.128.087002,Law2024}. It is particularly prominent in narrow-band systems, as magic-angle twisted or biased multi-layer graphene~\cite{Cao2018,Hao2021,Park2021,Zhou2021,Zhou2022,Andrei2020}, and
in nonlinear effects~\cite{Gao2023,Wang2023,Moore2025}.


Despite its fundamental role in solids, the QGT is measured essentially in engineered platforms emulating electronic bands
~\cite{10.1093/nsr/nwz193,PhysRevLett.122.210401,Zheng_2022,PhysRevResearch.5.L032016,PhysRevResearch.6.L022020,guillot2025,guillot2025nonabelian,Gianfrate:2020aa}.
In contrast, experiments in solids remain exceedingly rare, with a few recent realizations in angle-resolved photoemission spectroscopy~\cite{Kang_2024,Kim2025}.
In this context, quasiparticle interference (QPI) maps, imaged with atomic resolution by scanning tunneling microscopy (STM), have emerged as a powerful alternative for probing electronic band structures in two-dimensional materials.
They consist of standing-wave modulations of the local density of states (LDOS) due to elastic scattering on atomic-scale defects.
QPI maps enable the reconstructions of band-structure features in various systems, such as their low-energy bands \cite{petersenDirectImagingTwodimensional1998a,simon2011fourier,Avraham2018,Mallet2012} and \textit{global} topological properties~\cite{Roushan2009,Mallet2012,Fang2013,Dutreix2016Rhombohedral,PhysRevB.96.195207,Dutreix_2019,PhysRevLett.125.116804,PhysRevLett.125.176404,Dutreix_2021,Zhang_2021,Goft2023,Guan_2024,PhysRevLett.133.036204,Abulafia2025,Engstrom_2025,mesple2025,Bercioux2026}.
%
The latter are observable with conventional unpolarized STM tips, which lack orbital resolution.
However, resolving the \textit{local} quantum geometry requires access to the orbital character of wave functions and remains an experimental challenge.

In this Letter, we introduce a tomography method to reconstruct the density matrix of electronic states from QPI maps obtained with unpolarized local probes.
We demonstrate this method for two-orbital models with honeycomb structure, focusing on the low-energy band structure near the Brillouin zone (BZ) corners.
For a single on-site impurity, backscattering between time-reversed constant energy contours (CECs) maps the density matrix of the pristine system through distinct orbital contributions.
Remarkably, we find that the full density matrix can be extracted from unpolarized QPI maps by exploiting the orthogonality of the point-group representations of the scattering states. 
This tomography further enables us to reconstruct the QGT, providing a practical method to access the full band structure from unpolarized LDOS measurements.
%


\begin{figure}[t]
    \includegraphics[width=\columnwidth, trim = 0cm 0cm 0cm 0cm, clip]{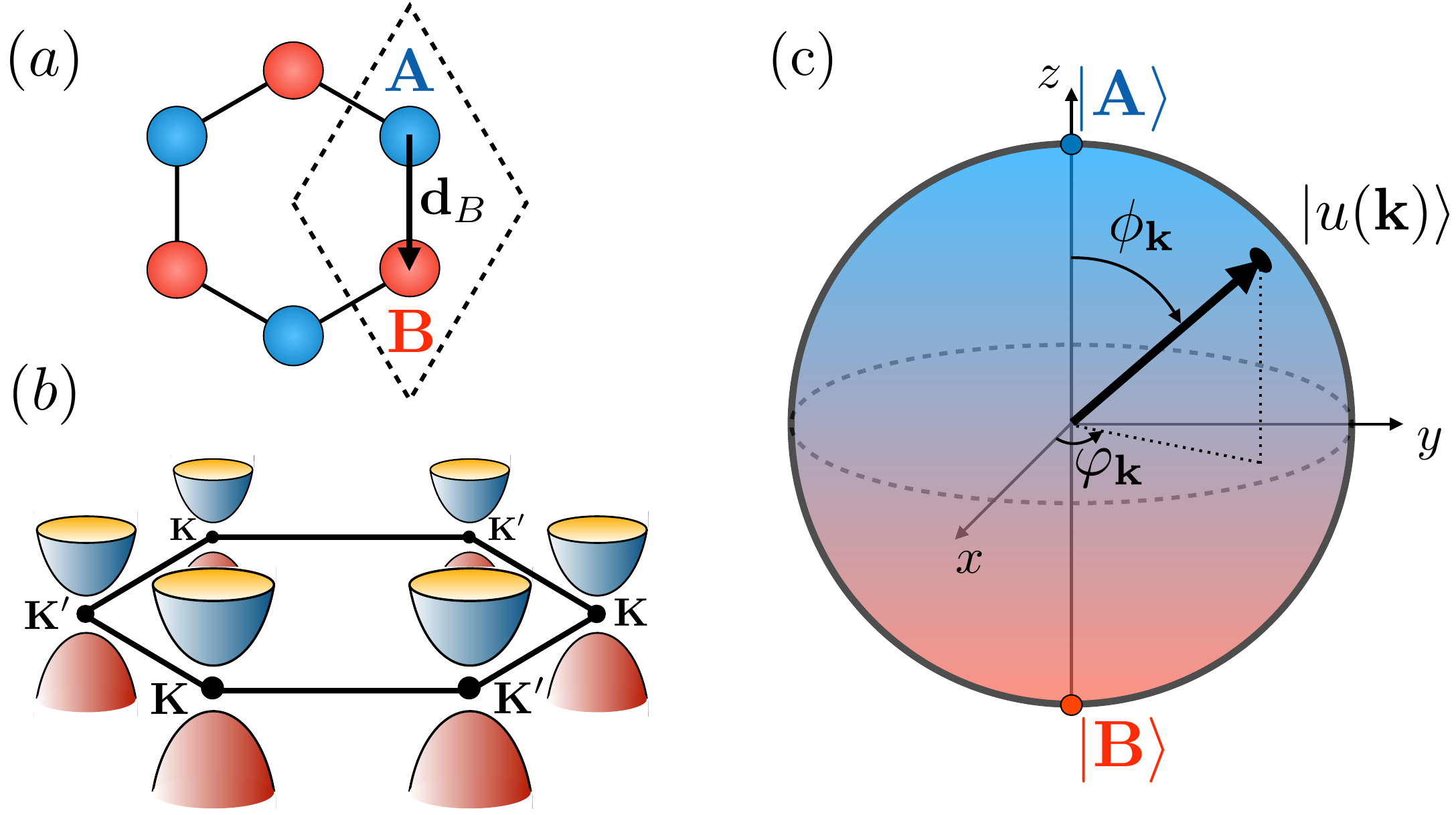}
    \caption{
    (a) Honeycomb lattice with two orbitals A and B per unit cell and relative position $\mathbf{d}_B$. 
    (b) Low-energy spectrum  with circular CECs at the BZ corners $\mathbf{K}$ and $\mathbf{K'}$.
    (c) Bloch sphere representation of the electron state $|u(\mathbf{k})\rangle$. The north (south) pole maps the orbital-polarized state $|A\rangle$ ($|B\rangle$).
    }
    \label{Lattice}
\end{figure}


\textit{Low-energy model} --- 
We consider a honeycomb lattice with time-reversal symmetry, where the low-energy band structure implies two spinless orbitals, A and B. 
These orbitals are located at different atomic sites in the unit cell, corresponding specifically to the two $2b$ Wyckoff positions [Fig.~\ref{Lattice}(a)].
The twofold ($C_2$) and threefold ($C_3$) symmetries enforce a two-dimensional irreducible representation of the little group at the BZ corners $\mathbf{K}$ and $\mathbf{K'}$. This ensures band degeneracies at these high-symmetry points, with circular CEC to leading order. Throughout this work, we relax $C_2$ symmetry, which may open a gap while preserving circular CECs in the valleys $\mathbf{K}$ and $\mathbf{K'}$ [Fig.~\ref{Lattice}(b)]. This two-band model captures the low-energy physics of various graphene heterostructures, such as monolayer graphene, gapped monolayer graphene on SiC or h-BN \cite{zhou2007substrate,Hunt2013,PhysRevLett.115.136802,chaves2020bandgap}, as well as spinless bands in transition metal dichalcogenides/diselenides \cite{Goerbig_2014}.

The low-energy band structure is generically described by a $2\times2$ Hamiltonian matrix $\hat{H}(\mathbf{k})$, where $\mathbf{k}$ is the momentum. We focus here on the conduction band and denote the corresponding normalized eigenstates as $|u(\mathbf{k})\rangle$. This pure state is characterized by the (projector) density matrix $\hat{\rho}(\mathbf{k})=|u(\mathbf{k})\rangle\langle u(\mathbf{k})|$. In the basis of the orbital states $|A\rangle$ and $|B\rangle$, it generically reads
\begin{align}\label{DM}
    \hat{\rho}(\mathbf{k})
    &=
    \frac{1}{2}
    \left(
    \begin{array}{cc}
    1+\cos(\phi_{\mathbf{k}}) & \sin(\phi_{\mathbf{k}})e^{-i\varphi_{\mathbf{k}}} \\
\sin(\phi_{\mathbf{k}})e^{i\varphi_{\mathbf{k}}} & 1-\cos(\phi_{\mathbf{k}})
    \end{array}
    \right)~.
\end{align}
The spherical angles $(\phi_{\mathbf{k}}, \varphi_{\mathbf{k}})$ parametrize the eigenstate representation on the Bloch sphere [Fig.~\ref{Lattice}(c)]. The polar angle $\phi_{\mathbf{k}}$ determines the orbital populations $P_\alpha = \langle \alpha|\hat{\rho}|\alpha\rangle$ for $\alpha \in \{A,B\}$, while the azimuthal angle $\varphi_{\mathbf{k}}$ defines the phase of the quantum coherence $C_{AB} = \langle A|\hat{\rho}|B\rangle = C_{BA}^*$.


\textit{Unpolarized QPI maps} ---
We focus on the QPI map surrounding a single impurity, as is typical in STM samples with dilute disorder.
We model it by a $\delta$-localized potential $\hat{V}\delta(\mathbf{r})$, where $\hat{V}$ is a $2\times2$ matrix encoding the orbital structure of the defect.
%
%
%
We treat elastic scattering in a $\hat{T}$-matrix approach \cite{SM}, which includes all scattering orders and captures realistic resonant impurities \cite{katsnelson2020,Kaasbjerg2020}.

\begin{figure}[t]
    \centering
    \includegraphics[scale=0.25, trim = 0cm 0cm 0cm 0cm, clip]{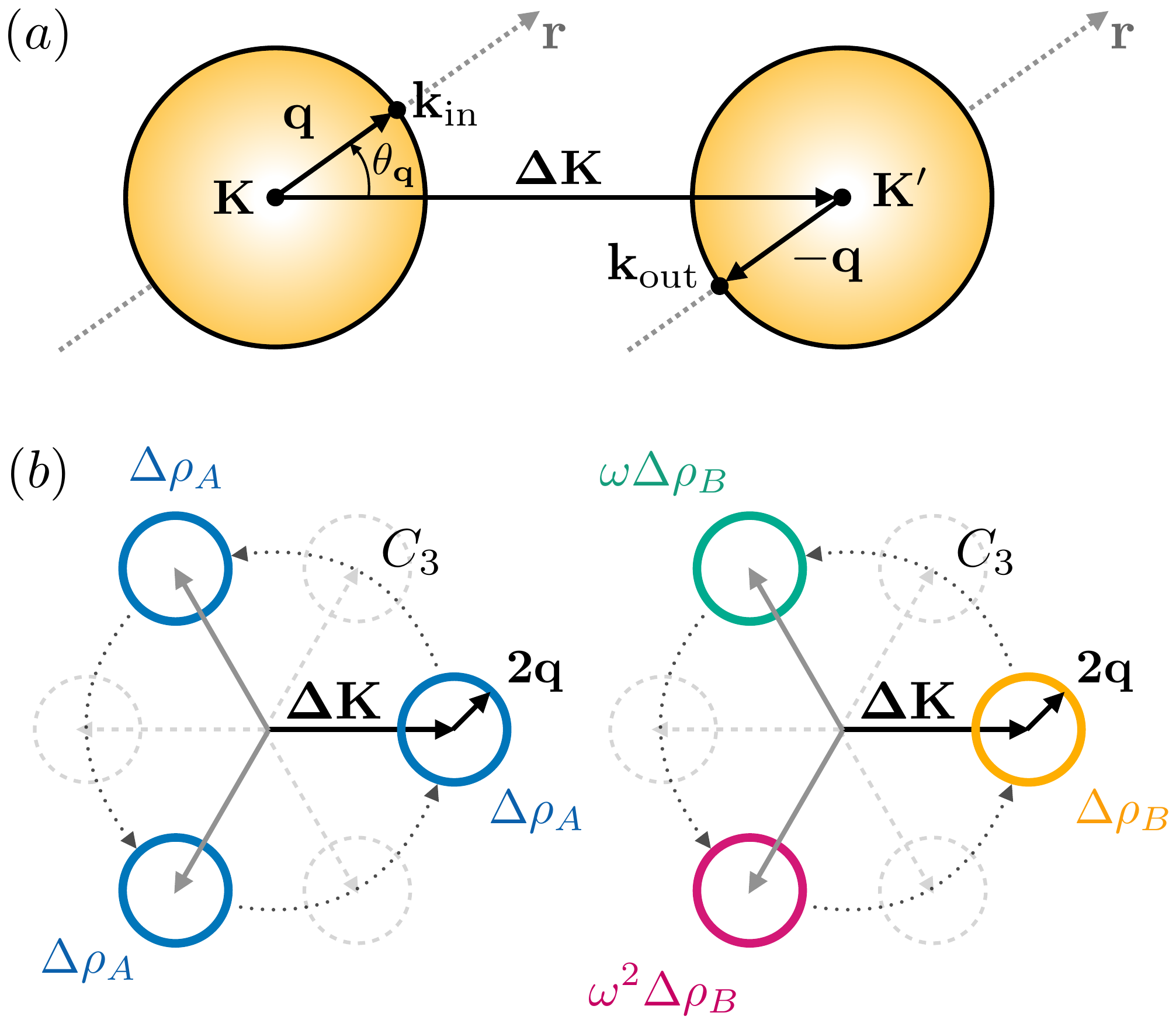}
    \caption{
    (a) Elastic backscattering process coupling the pair of stationary points $\mathbf{k}_{\rm in} = \mathbf{K} + \mathbf{q}$ and $\mathbf{k}_{\rm out} = \mathbf{K'} - \mathbf{q}$ along two CECs.
    (b) Orbital contributions in Fourier space showing $2q$-radius rings at three intervalley vectors $\mathbf{\Delta K}$ related by $C_3$ symmetry: $\Delta\rho_{A}$ remains unchanged under $C_3$ rotation, while $\Delta\rho_B$ picks a phase factor $\omega=\exp(i\Phi_B)$.
    }
    \label{BackscatteringQPI}
\end{figure}

To evaluate the resulting LDOS modulations $\Delta\rho(\mathbf{r},E)$ at the STM tip position $\mathbf{r}$ and energy $E$, we resort to the stationary phase approximation \cite{Roth1966,Lounis2011,Liu2012}.
It captures the dominant scattering contributions from points along CECs where the group velocity is parallel or antiparallel to $\mathbf{r}$.
For circular CECs, the leading contributions arise from backscattering between the stationary points $\mathbf{k}_{\rm in} = \mathbf{K} + \mathbf{q}$ and $\mathbf{k}_{\rm out} = \mathbf{K'} - \mathbf{q}$ [Fig.~\ref{BackscatteringQPI}(a)].
The resulting LDOS modulations exhibit universal $2q$ wavevector oscillations with algebraic decay at large distances:
\begin{align}\label{LDOS_SPA}
	\Delta \rho (\mathbf{r},E) 
	&\propto
    \text{Re} 
	\Big[
    \frac{e^{i(\mathbf{\Delta K}-2\mathbf{q})\cdot\mathbf{r}}}{r} 
    \text{Tr}
    [
    \hat{\rho}(\mathbf{k}_{\rm out})
    \hat{T}
    \hat{\rho}(\mathbf{k}_{\rm in})
    ]
    \Big]
    ~,
\end{align}
where $\mathbf{\Delta K} = \mathbf{K'} - \mathbf{K}$ is the intervalley vector, and we omit an overall (dimensional) prefactor for clarity \cite{SM}.
The trace $\Tr$ averages over the orbital states $|A\rangle$ and $|B\rangle$, such that $\Delta \rho$ is directly observable through conventional unpolarized LDOS measurements.
%

In STM experiments, such LDOS modulations are usually analyzed in Fourier space, where they manifest as high-intensity rings centered at the intervalley vector $\mathbf{\Delta K}$ \cite{petersenDirectImagingTwodimensional1998a,simon2011fourier,Avraham2018,Mallet2012}.
The rings mirror the circular CECs involved in backscattering, which enables the reconstruction of low-energy bands from a series of QPI maps at different energies.
These rings may also exhibit intensity extinctions for specific orientations, due to selection rules informative over the chirality or topological obstruction of the quasiparticle wave functions \cite{mesple2025,Roushan2009,Mallet2012,Fang2013,Bercioux2026}.
These features are directly visualized in conventional QPI maps, where the unpolarized STM tips lack orbital resolution and average over distinct orbital contributions.
We now show that these orbital contributions can, in fact, be extracted from unpolarized QPI maps to reveal the quasiparticle density matrix.

\textit{Orbital contributions} ---
As captured by the trace in Eq.~(\ref{LDOS_SPA}), the total LDOS modulations can be decomposed as $\Delta \rho = \Delta \rho_A + \Delta \rho_B$.
The Fourier transform of the orbital contribution $\Delta\rho_{\alpha}$ in the vicinity of the intervalley vector $\mathbf{\Delta K}$ verifies~\cite{SM}
\begin{align}\label{QPI_LDOS}
    \Delta\rho_{\alpha}(\mathbf{\Delta K},\mathbf{Q})
    &\propto\frac{\mathcal{F}_{\alpha}(\mathbf{\Delta K},\mathbf{Q})}{\sqrt{|Q^2-(2q)^2}|}
    ~,
\end{align}
where $\mathbf{Q}$ is a small scattering vector from $\mathbf{\Delta K}$, $q$ the radius of the CEC at energy $E$, and we omit identical prefactors encoding the potential strength, group velocities, and CEC curvature at the stationary points.
The effective divergency in the denominator is removed in practice by assuming a finite quasiparticle lifetime \cite{petersenDirectImagingTwodimensional1998a, simon2011fourier,Avraham2018,Mallet2012}.
As for unpolarized QPI maps~\cite{petersenDirectImagingTwodimensional1998a,simon2011fourier,Avraham2018,Mallet2012}, this produces a high-intensity $2q$-radius ring at the intervally vector $\mathbf{\Delta K}$ [Fig.~\ref{BackscatteringQPI}(b)]. 

While both orbital contributions exhibit $2q$-radius rings, they are generally modulated by different orbital factors $\mathcal{F}_{\alpha}$.
The orbital factors are determined by the multi-orbital structure of the scattering states and the impurity. 
We find that for a single on-site potential, leading to the $\mathbf{k}$-independent matrix $\hat{V} = V|A\rangle\langle A|$ in momentum space, the orbital factor takes the explicit form~\cite{SM}:
\begin{align}\label{OrbFact1}
    \mathcal{F}_{\alpha}(\mathbf{\Delta K}, \mathbf{Q})
    &=
    \langle \alpha |
    \hat{\rho}(\mathbf{k}_{\rm out})
    |A\rangle
    \langle A |
    \hat{\rho}(\mathbf{k}_{\rm in})
    | \alpha \rangle
    ~.
\end{align}
Thus, the on-site impurity directly maps the density matrix populations (coherences) of the incoming and outgoing scattering states into the orbital modulation $\mathcal{F}_{A(B)}$ of the $2q$-radius rings.
Reconstructing the density matrix therefore requires the resolution of each orbital contribution individually.

We propose to extract each orbital contribution $\Delta\rho_\alpha$ from conventional unpolarized QPI maps by exploiting the point group representations of the scattering states.
In particular, backscattering occurs between CECs centered at the BZ corners. 
These points are invariant under $C_3$ symmetry and satisfy $C_3^n\mathbf{K} = \mathbf{K}+\mathbf{G}$, where $\mathbf{G}$ is a reciprocal lattice vector and $n\in\{0,1,2\}$.
This implies that the orbital contribution at the intervalley vector $C_3^n\mathbf{\Delta K}$ relates to that at the intervalley vector $\mathbf{\Delta K}$ as
\begin{align}\label{Irreps}
    \Delta\rho_{\alpha}(C_3^n\mathbf{\Delta K}, \mathbf{Q})
    &=
    e^{in\Phi_{\alpha}} \Delta\rho_{\alpha}(\mathbf{\Delta K}, \mathbf{Q})
    ~,
\end{align}
where $\Phi_{\alpha}=C_3\mathbf{\Delta K}\cdot(C_3\mathbf{d}_{\alpha}-\mathbf{d}_{\alpha})$, $\mathbf{d}_\alpha$ is the Wyckoff position of orbital $\alpha$ (with respect to that of the scattering potential), and the vector $C_3\mathbf{d}_{\alpha}-\mathbf{d}_{\alpha}$ matches a Bravais lattice vector~\cite{SM}.
Each orbital contribution must return to itself after three successive symmetry operations, since $C_3^3=I$.
The phase $\Phi_\alpha$ thus corresponds to the $3$-rd roots of unity and determines the one-dimensional irreducible representation (irrep) of the cyclic point group $C_{3}$. 
The total LDOS modulations $\Delta\rho$ at intervalley vectors, as measured by unpolarized STM tips, therefore satisfy
\begin{align}\label{IrrepsProjection}
    \left(
    \begin{array}{c}
    \Delta\rho(\mathbf{\Delta K}) \\
    \Delta\rho(C_3\mathbf{\Delta K}) \\
    \Delta\rho(C_3^2\mathbf{\Delta K}) \\
    \end{array}
    \right)
    &=
    \Delta\rho_{A}(\mathbf{\Delta K})
    |\Gamma_A\rangle
    +
    \Delta\rho_{B}(\mathbf{\Delta K})
    |\Gamma_B\rangle
    ~,
\end{align}
where $\langle\Gamma_\alpha|=(1,e^{i\Phi_{\alpha}},e^{i2\Phi_{\alpha}})$ is the irrep vector determining how the orbital contribution $\Delta\rho_\alpha$ transforms under $C_3$. 
For clarity, we omit the $\mathbf{Q}$-dependence in Eq.~(\ref{IrrepsProjection}).

For orbitals $A$ at Wyckoff position $\mathbf{d}_A=0$, Eq.~(\ref{Irreps}) shows that contribution $\Delta\rho_A$ always transforms under the trivial irrep [Fig.~\ref{LDOS_SPA}(b)].
Thus, the corresponding irrep vector satisfies $\langle\Gamma_A|=(1,1,1)$.
In contrast, the Wyckoff position of orbitals $B$ satisfies $\mathbf{d}_B\neq0$.
Thus, contribution $\Delta\rho_{B}$ necessarily transforms under a non-trivial irrep ($\Phi_B=\pm2\pi/3$) for backscattering between non-equivalent valleys ($\mathbf{\Delta K}\neq\mathbf{G}$).
Since the two orbital contributions transform under different irreps, the corresponding vectors $|\Gamma_A\rangle$ and $|\Gamma_B\rangle$ are necessarily orthogonal, as guaranteed by the Wonderful Orthogonality Theorem in group theory \cite{dresselhaus2007group}.
It is then remarkable that each orbital contribution $\Delta\rho_{\alpha}$ can be extracted individually by projecting the total unpolarized signal $\Delta\rho$ onto the corresponding irrep vectors.
%
%

\begin{figure*}[t]
    \centering
    \includegraphics[width=1.\linewidth]{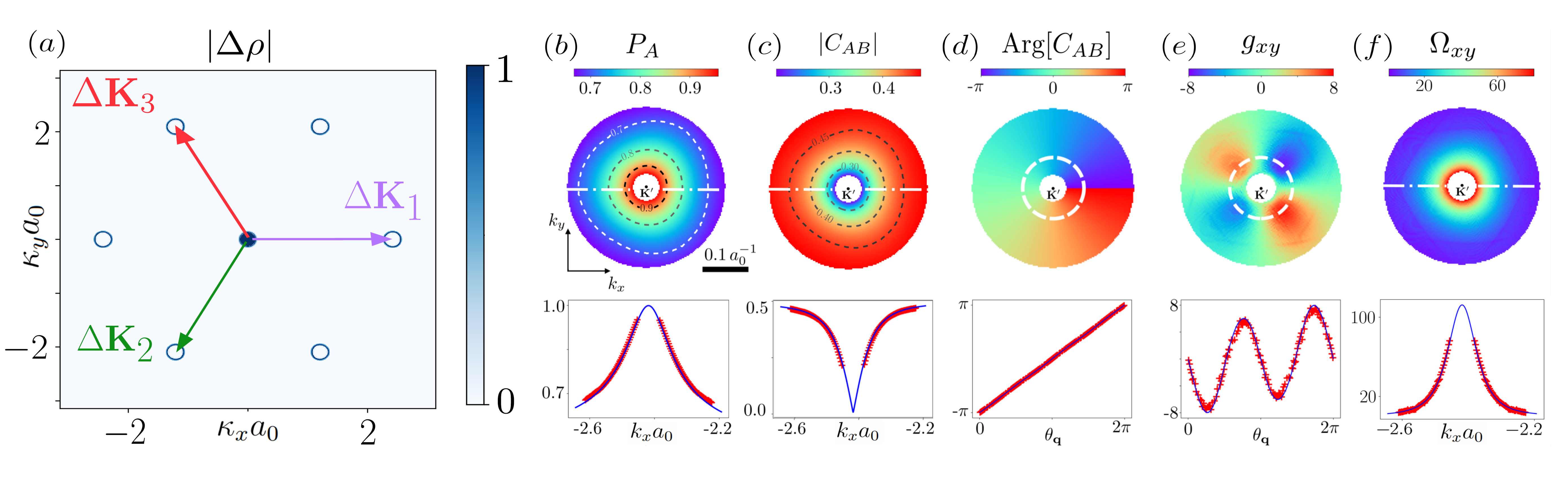}
    \caption{(a) Unpolarized QPI map showing the LDOS intensity $|\Delta\rho|$ in Fourier space at energy $E=0.15t$. The vectors $\boldsymbol{\Delta K}_{1,2,3}$ point toward intervalley backscattering intensity rings related by $C_3$ symmetry.
    (b-f) 2D maps of the density matrix and geometric tensor reconstructed in valley $\mathbf{K'}$ from intervalley backscattering rings as shown in panel (a): (b) population $P_{A}$, (c) coherence modulus $|C_{BA}|$, (d) coherence phase $\text{Arg}[C_{BA}]$, (e) quantum metric $g_{xy}$, and (f) Berry curvature $\Omega_{xy}$. 
    The components $g_{xx}$, $g_{yy}$, $\Omega_{xx}$ and $\Omega_{yy}$ are shown in Ref.~\cite{SM}.
    The bottom row shows cuts of the QPI-reconstructed quantities (red crosses) along the white dashed lines in the 2D maps on top. They compare to their theoretical counterparts (blue solid lines) evaluated from the bare band structure (\ref{BH}).
    Numerical parameters are $N=1$, $\Delta_0 = 0.1t$, $V = 0.4t$, $\eta = 0.002t$.
    }
    \label{fig:ReconstructionPopCohQGT}
\end{figure*}


\textit{Electron state tomography} ---
Backscattering between nearest CECs couples TRS conjugate stationary points, satisfying $\mathbf{k}_{\rm out} = -\mathbf{k}_{\rm in} + \mathbf{G}$, where $\mathbf{G}$ is a reciprocal lattice vector.
Under TRS, the populations and coherences satisfy $P_\alpha(\mathbf{k}) = P_\alpha(-\mathbf{k})$ and $C_{AB}(\mathbf{k}) = C_{BA}(-\mathbf{k})$, respectively, thus leading to the orbital factors
\begin{align}
    \mathcal{F}_{A}(\mathbf{\Delta K}, \mathbf{Q})
    &=
    P_{A}^2(\mathbf{k}_{\rm out})
    ~,
    \\
    \mathcal{F}_{B}(\mathbf{\Delta K}, \mathbf{Q})
    &=
    C_{BA}^2(\mathbf{k}_{\rm out}) \,  e^{i\mathbf{G}\cdot\mathbf{d}_{B}}
    ~.
\end{align}
Remarkably, these orbital factors respectively map (the squares of) the population $P_{A}$ and coherence $C_{BA}$ of the density matrix of the outgoing state $|u(\mathbf{k}_{\rm out})\rangle$ --- or the incoming state $|u(\mathbf{k}_{\rm in})\rangle$ by TRS. Since the density matrix is a projector, the populations and coherences are not independent: $|C_{AB}|^2=P_AP_B$. We can thus reconstruct the outgoing-state density matrix in a way that is independent of the potential strength $V$:
\begin{align}\label{eq:PopCohQPI}
    P_{\alpha}
    &=
    \frac{|\Delta\rho_{\alpha}|}{\sum_{\beta}|\Delta\rho_{\beta}|}
    ~~~\text{and}~~~
    2\varphi+\mathbf{G}\cdot\mathbf{d}_B
    =
   \arg \frac{\Delta\rho_{B}}{\Delta\rho_{A}}
    ~,
\end{align}
where we omit the wave-vector dependence in $P_\alpha$, $\varphi$, and $\Delta\rho_{\alpha}$ for clarity~\cite{SM}.
%
%
This shows how the resolution of the orbital contributions enables the tomography of electron states along circular CECs related by TRS.

From a series of QPI maps at different energies, which probes the CECs of different radia $q$, we can obtain a two-dimensional map characterizing the electron states in the vicinity of the BZ corners.
Combined with standard spectroscopic STM measurements of the dispersion relation \cite{petersenDirectImagingTwodimensional1998a,simon2011fourier,Mallet2012,Avraham2018,mesple2025}, this leads to complete knowledge of the band structure --- energy bands and electron density matrix.
This shows that we can further access the QGT capturing the local quantum geometry of the Hilbert space nearby the high-symmetry points:
$\mathcal{Q}_{xy}
    = \Tr[(\partial_x\hat{\rho})(1-\hat{\rho})(\partial_y\hat{\rho})]$,
where $g_{xy}=\Real[\mathcal{Q}_{xy}]$ defines to the quantum metric and $\Omega_{xy}=-2\Imag[\mathcal{Q}_{xy}]$ the Berry curvature.

\textit{Application to numerical QPI maps} ---
To benchmark the tomography method, we apply it to QPI maps obtained numerically from tight-binding models on a honeycomb lattice and verify that it reproduces the correct density matrix.
The generic Bloch Hamiltonian is ($\hbar=1$)
\begin{align}\label{BH}
    \hat{H}(\mathbf{k})
    &=
    \left(
    \begin{array}{cc}
    \Delta(\mathbf{k}) & {tf(\mathbf{k})} \\
    {tf^*(\mathbf{k})} & -\Delta(\mathbf{k})
    \end{array}
    \right)~,
\end{align}
where $\Delta(\mathbf{k})$ represents a staggered sublattice potential, $t$ denotes the nearest-neighbor hopping amplitude, and $f(\mathbf{k})=e^{i\mathbf{k}\cdot \mathbf{d}_{B}}(1+e^{i\mathbf{k}\cdot \mathbf{a}_{1}}+e^{i\mathbf{k}\cdot \mathbf{a}_{2}})$ with $\mathbf{a}_{1}=a_0(\sqrt{3},3)/2$ and $\mathbf{a}_{2}=a_0(-\sqrt{3},3)/2$ the Bravais vectors, and $a_0$ the lattice constant.
%
The mass term $\Delta(\mathbf{k})$ enables the interpolation between nodal semimetals and direct gap semiconductors
\cite{zhou2007substrate,Hunt2013,PhysRevLett.115.136802,chaves2020bandgap,Goerbig_2014}.

We first present numerical results for $\Delta(\mathbf{k})=\Delta_0$, which captures the low-energy physics of massive Dirac quasiparticles.
We compute the unpolarized QPI maps directly in Fourier space via a numerical $\hat{T}$-matrix approach, based on the retarded bare Green function $\hat{G}_0(\mathbf{k},E)=[E+i\eta-\hat{H}(\mathbf{k})]^{-1}$, where $\eta$ is the inverse quasiparticle lifetime [Fig.~\ref{fig:ReconstructionPopCohQGT}(a)].
We extract the orbital contributions $\Delta\rho_\alpha$ in the vicinity of the intervalley vectors by exploiting their symmetry group representation.
Applying the prescription in Eq.~(\ref{eq:PopCohQPI}) to each pixel of the intervalley $2q$-radius ring in $\Delta\rho_\alpha$, we obtain the populations and coherences of the scattering states along the outgoing CEC.
We repeat this procedure varying the energy $E$ to map the $\mathbf{k}$-dependence of the reconstructed density matrix (see Ref.~\cite{SM} for methodological details).

Figures~\ref{fig:ReconstructionPopCohQGT}(b-f) show 2D maps of the QPI-reconstructed density matrix $\hat{\rho}(\mathbf{k})$ in valley $\mathbf{K'}$.
These maps exhibit several qualitative features directly expected from the band structure in Eq.~(\ref{BH}).
The reconstructed population $P_A$ approaches 1 near the conduction band minimum $\Delta_0$ at $\mathbf{K'}$, where the electron states are entirely polarized on orbital $A$ [Fig.~\ref{fig:ReconstructionPopCohQGT}(b)].
Consistently, the reconstructed coherence $C_{BA}$, which quantifies the overlap between the orbital states $A$ and $B$, vanishes at $\mathbf{K'}$ [Fig.~\ref{fig:ReconstructionPopCohQGT}(c)].
Also the coherence phase seems to vary linearly around $\mathbf{K'}$, as expected from the pseudospin-momentum locking at low energy, and leads to a global pseudospin winding of $2\pi$ [Fig.~\ref{fig:ReconstructionPopCohQGT}(d)].
From these maps, we can further reconstruct the quantum metric $g_{xy}$ [Figs.~\ref{fig:ReconstructionPopCohQGT}(e)] and Berry curvature $\Omega_{xy}$ [Figs.~\ref{fig:ReconstructionPopCohQGT}(f)] in valley $\mathbf{K'}$.
%
%
To be more quantitative, we compare in the bottom panels in Fig.~\ref{fig:ReconstructionPopCohQGT} these QPI-reconstructed quantities with their counterparts directly obtained from the band structure in Eq.~(\ref{BH}).
The excellent agreement between the two confirms the reliability of the tomography method for probing electron states.

\textit{Range of validity} --- 
The tomography method we introduced above should also apply to any scattering potential strength and to various two-band systems with time-reversal and hexagonal symmetries.
In Ref.~\cite{SM}, we further test numerically the method validity for graphene's massless Dirac fermions where $\Delta_0=0$, for a $\mathbf{k}$-dependent mass term $\Delta(\mathbf{k})$, arising from nonequivalent next-nearest neighbor hopping amplitudes, for both weak and strong scattering potentials $V$, and for the inter-orbital coupling $(tf(\mathbf{k}))^2$ instead of $tf(\mathbf{k})$ in Hamiltonian (\ref{BH}).
For all these different systems, the QPI-based tomography shows excellent agreement with the density matrix of the bare band structure.
We also find numerically that the method works surprisingly well for CECs with trigonal warping at higher energies, suggesting that its applicability extends beyond the isotropic assumption of circular CECs~\cite{SM}.
In fact, the method is here limited by the inverse life time $\eta$, which determines the CEC width.
When this width becomes comparable to the CEC radius $q$, backscattering can no longer be resolved.
This occurs near the high-symmetry points at low energy, leading to the blank areas observed at the center of the 2D maps in Fig.~\ref{fig:ReconstructionPopCohQGT}.

\textit{Conclusion} --- 
In this Letter, we introduced a tomography method to reconstruct the density matrix of electronic states from QPI maps in Fourier space, as obtained with conventional, unpolarized STM tips.
For two-orbital models with honeycomb structure, we showed that intervalley backscattering on single on-site impurities leads to complete reconstruction of the band structure --- energy bands and electron states.
The density matrix further provides direct knowledge of the QGT characterizing the quantum geometry of the electron states in momentum space.
This establishes impurities as tomographic probes for STM experiments using unpolarized tips, with direct applications to various van der Walls hexagonal heterotructures, whose low-energy quantum geometry is governed by a two-orbital basis.

While completing our work, we became aware of a recent work by Zhang \textit{et al.}~\cite{SHZhang2026}, who proposed a real-space reconstruction of the QGT from the QPI maps of a spin system with magnetic impurities and observable with spin-polarized STM tips.
In contrast, our tomography approach shows that one can resolve the density matrix, for a (spinless) two-orbital system, and in Fourier space.
This relies crucially on disentangling the distinct orbital contributions from unpolarized LDOS measurements, which is achieved via symmetry group analysis.
This is the knowledge of the density matrix that further leads to the QGT in our case.
By focusing on unpolarized local probes, our method is thus experimentally relevant not only for electronic bands but also for engineered lattices, where unpolarized LDOS measurements are routinely accessible~\cite{Bellec_2013,Bellec_2014,Amo_2014,Torrent_2012,Rechtsman_2013,tarruell_2012,jotzu_2014}.
Finally, we emphasize that our approach complements that of Zhang \textit{et al.}~\cite{SHZhang2026}, by paving the way for STM investigations of the QGT in more complex materials with both spin and orbital degrees of freedom.

\textit{Acknowledgements.} --
We acknowledge support from the project TopoMat (ANR-23-CE30-0029) funded by the French Research National Agency.

\bibliography{references}

\end{document}